\documentclass[prb,aps,showpacs,twocolumn]{revtex4}

\usepackage{amsmath,amsfonts}

\usepackage[utf8]{inputenc}
\usepackage{bm}

\usepackage{graphicx}
\usepackage{subfig}

\usepackage[squaren]{SIunits}

\usepackage{hyperref}

\DeclareMathOperator{\sgn}{sign}
\newcommand{\signum}[1]{ \sgn \left( #1 \right) }
\newcommand{\order}[2]{ O(#1^{#2}) }

\newcommand*{\up}{ \uparrow }
\newcommand*{\down}{ \downarrow }

\newcommand{\pdiff}[2]{ \ensuremath{ \frac{\partial #1}{\partial #2} } }
\newcommand{\abs}[1]{ \ensuremath{\left| #1 \right|} }
\newcommand*{\im}{\ensuremath{\mathrm{i}}}

\newcommand{\trace}[1]{\ensuremath{\mathrm{Tr}\left[#1\right]}}

\begin{document}

\title{Crossed Andreev reflection versus electron transfer in graphene
  nanoribbons }

\author{H\aa vard Haugen}
\author{Daniel Huertas--Hernando}
\author{Arne Brataas}
\affiliation{Department of Physics, 
  Norwegian University of Science and Technology,
  N-7491 Trondheim, Norway}

\author{Xavier Waintal}
\affiliation{SPSMS-INAC-CEA,
  17 rue des Martyrs,
  38054 Grenoble CEDEX 9,
  France}

\date{\today}

\begin{abstract}
  We investigate the transport properties of three-terminal graphene
  devices, where one terminal is superconducting and two are normal
  metals. The terminals are connected by nanoribbons. Electron
  transfer (ET) and crossed Andreev reflection (CAR) are identified
  via the non-local signal between the two normal
  terminals. Analytical expressions for ET and CAR in symmetric
  devices are found. We compute ET and CAR numerically for asymmetric
  devices. ET dominates CAR in symmetric devices, but CAR can dominate
  ET in asymmetric devices, where only the zero-energy modes of the
  zigzag nanoribbons contribute to the transport.
\end{abstract}

\maketitle

\section{Introduction}

Graphene, a two-dimensional honeycomb lattice of carbon atoms, has
recently been experimentally realized.\cite{Novoselov2004sci306,
  Zhang2005n438, Novoselov2005n438} It exhibits intriguing electron
transport properties such as a very high
mobility,\cite{Novoselov2004sci306, Zhang2005n438, Novoselov2005n438}
gate-voltage tunable electron doping,\cite{Novoselov2004sci306}
anomalous quantum Hall effect,\cite{Zhang2005n438} Klein
tunneling,\cite{Katsnelson2006nphys} ``relativistic'' Dirac-like
linear energy-momentum
dispersion,\cite{semenoff:prl:v53:p2449:y1984:honeycomb_dirac_equation}
and possible integration with other adatoms and electrical
contacts.\cite{geim:nmat:v6:p183:y2007:rise_of_grapene,
  castroneto:rmp:v81:p109:y2009:electronic_prop_graphene} Graphene can
be contacted to superconductors and a supercurrent in graphene
Josephson junctions has been
measured.\cite{heersche:ssc:v143:p72:y2007:supercondgraphene,
  du:prb:v77:p184507:y2007:multiple:andreev:reflections:SNS:junctions,
  kessler:prl:v104:p047001:y2010:superconducting:metal:clusters:on:graphene}

Non-local transport in three-terminal devices with one superconducting
lead and two normal metals has been extensively studied, both
theoretically\cite{ byers:prl:v74:p306:y1995:two_contact_tunneling,
  deutscher:apl:v76:p487:y2000:SF_point_contacts_Andreev_reflection,
  falci:epl:v54:p255:y2001:correlated_tunneling_superconductor,
  bignon:epl:v67:p110:y2004:current_current_corr,
  morten:prb:v74:p214510:y2006:circuit_theory_CAR,
  morten:prb:v78:p224515:y2008:fcs} %
and experimentally.\cite{
  beckmann:prl:v93:p197003:y2004:evidence_for_CAR_in_SF_structures,
  russo:prl:v95:p027002:y2005:bias_dep_nonlocal_AR,
  cadden-zimansky:prl:v97:p237003:y2006:nonlocal_NSN_systems,
  kleine:epl:v87:p27011:y2009:contact_resistance_dependence_of_CAR} At
energies lower than the superconducting gap, the current in one normal
terminal caused by a voltage applied between another normal terminal
and the superconductor is governed by a competition between electron
transfer (ET) and crossed Andreev reflection (CAR). ET is the emission
of an electron from one normal metal terminal to another normal metal
terminal, possibly after interacting with the superconductor. In CAR,
an electron from one normal terminal enters the superconductor
together with an electron from a second normal terminal or,
equivalently, an electron is emitted into one normal terminal while a
hole is emitted into another normal terminal. This process creates a
spatially entangled electron-hole pair which has a fundamental
interest and can be used as an
entangler.\cite{recher:prb:v63:p165314:y2003:andreev-entangled-electrons,
  prb:v66:p161320:y2002:bell-inequalities-solid-state,
  prl:v91:p267003:y2003:coulomb-blockade-BCS-spin-entangled-electrons}
The relative magnitude of ET and CAR can be tuned by using
ferromagnetic
contacts,\cite{falci:epl:v54:p255:y2001:correlated_tunneling_superconductor,
  melin:jphys-condmat:v13:p6445:y2001:FSF} but our focus here is on
their intrinsic relative value when normal metals are used. The ET and
CAR processes contribute with opposite signs to the non-local current.
Experimentally, it has been measured that ET often dominates CAR, but
at finite bias voltage a CAR dominated
signal\cite{russo:prl:v95:p027002:y2005:bias_dep_nonlocal_AR} was also
seen.  First theories in the lowest order tunneling limit predict that
ET and CAR exactly cancel each
other.\cite{falci:epl:v54:p255:y2001:correlated_tunneling_superconductor}
Also, relaxing the assumption of tunneling barriers by allowing
barriers of arbitrary strength in semi-classical N-S circuits, ET
generally dominates
CAR.\cite{morten:prb:v74:p214510:y2006:circuit_theory_CAR} Recent
theoretical suggestions to explain the experiment in
Ref.~\onlinecite{russo:prl:v95:p027002:y2005:bias_dep_nonlocal_AR} are
Coulomb interaction
effects,\cite{npys:v3:p455:y2007:yeyati_andreev_pairs_and_collective_excitations}
an external AC
bias,\cite{golubev:epl:v86:p37009:y2009:non-local-AR-under-ac-bias}
and quantum interference
effects.\cite{golubev:prl:v103:p067006:y2009:CAR-and-charge-imbalance}

Theoretically, graphene-superconductor junctions have been
investigated by several workers.
\cite{beenakker:prl:v97:p067007:y2006:specular_andreev,
  titov:prb:v74:p041401:y2006:josephson_effect_in_graphene,
  burset:prb:v77:p205425:y2008:microscopic_proximity_effect_GS} In
graphene there is an additional new quasi-particle to supercurrent
conversion process denoted specular Andreev reflection, where states
above and below the Dirac point are coupled by Andreev scattering
(inter-band
coupling).\cite{beenakker:prl:v97:p067007:y2006:specular_andreev} In
specular Andreev reflection, the holes emitted from the superconductor
no longer follow the parallel time-reversed path of the incoming
electron as they do in direct Andreev reflection, but are specularly
reflected at an angle which equals the angle of incidence. Although
fundamentally interesting, as it could enhance CAR
processes,\cite{benjamin:prb:v78:p235403:y2008:specular_CAR,
  cayssol:prl:v100:p147001:y2008:CAR_in_graphene_pn_transistor}
specular Andreev reflection is only visible in ultra clean and
homogeneous systems, since the Dirac point must be well-defined
throughout the region of interest, or the superconducting pair
potential $\Delta$ must be much larger than the typical variation in
the Dirac point. Also, it is necessary to control the doping such that
the Fermi energy is considerably smaller than the superconducting
gap. The small magnitude of the proximity induced superconducting gap
in graphene, $\Delta \approx \unit{0.1}{\milli\electronvolt
}$,\cite{heersche:ssc:v143:p72:y2007:supercondgraphene} means this
could only be realized in ultra small structures at very low doping
level in well controlled systems.

In this paper we investigate the influence of a superconducting
terminal on devices built from graphene zigzag ribbons. We are
interested in studying how ET and CAR depend on the features of the
nanoribbons \textit{e.g.} on their widths, number of terminals, and
relative angle where ribbons connected to various terminals
intersect. The choice of zigzag ribbons is the most relevant one, as
the boundary conditions for ribbons with generic boundaries have been
shown to reduce to the boundary conditions for zigzag terminations in
most
cases.\cite{akhmerov:prb:v77:p085423:y2008:boundary_zigzag_generic,
  rycerz:pssa:v205:p1281:y2008:nonequilibrium_valley_polarization}
Such nanoribbons have some unique electronic features, such as
supporting current carrying zero-modes localized along the
edges.\cite{fujita:jpsj:v65:p1920:y1996:localized_zigzag_edge_states,
  nakada:prb:v54:p17954:y1996:graphen_ribbon_edges,
  rycerz:pssa:v205:p1281:y2008:nonequilibrium_valley_polarization}
Also, for low energies, states carrying current in opposite directions
along the zigzag ribbon is associated with different eigenstates, and
there is an absence of backscattering due to the small overlap between
the states carrying current in opposite
directions.\cite{rainis:prb:v79:p115131:y2009:AR_in_graphene_ribbons}

The paper is organized as follows: In Sec.~\ref{sec:model} we define
our model and in Sec.~\ref{sec:scatt-matr-multi} we express the
scattering matrix in terms of the normal state scattering matrix. This
enables us to identify the ET and CAR contributions to the non-local
signal. Section~\ref{sec:symm-three-term} describes the properties of
symmetric three-terminal devices, and in Sec.~\ref{sec:car-dominance}
we present numerical results showing a dominance of CAR over ET in an
asymmetric device. Finally we conclude our paper in
Sec.~\ref{sec:conclusion}.

\section{Model and method}
\label{sec:model}

Our description of superconducting ribbons starts with the nearest
neighbor hopping Hamiltonian of graphene,
\begin{equation}
  \label{eq:graphene-hopping-hamiltonian}
  H 
  = 
  - \sum_{\langle i, j \rangle, \sigma} \gamma c_{i\sigma}^\dagger c_{j \sigma}
  - E_F \sum_{i, \sigma} c_{i \sigma}^\dagger c_{i \sigma}
  ,
\end{equation}
where $\gamma \approx \unit{2.7}{\electronvolt}$ is the nearest
neighbor hopping energy, \cite{wallace:pr:v71:p622:y1947,
  reich:prb:v66:p035412:y2002:tightbinding_descr_graphene,
  castroneto:rmp:v81:p109:y2009:electronic_prop_graphene} and $c_{i
  \sigma}^\dagger$ creates an electron with spin $\sigma$ at site $i$.
In the superconducting terminal, we assume a superconductor on top of
the graphene sheet which by proximity induces superconducting
properties in graphene. We consider spin singlet pairing described by
the BCS Hamiltonian $\hat{H}$.\cite{bardeen:pr:v108:p1175:y1957:BCS}
The superconducting pair potential $\Delta_i$ is local to site $i$ and
chosen to be real since we only have one superconductor. We write
$\hat{H}$
in Nambu form,
\begin{equation}
  \label{eq:Nambu}
  \hat{H}
  =
  \sum_{i,j}
  \Psi_j^\dagger
  \left[
  \begin{pmatrix}
    H_{i j} & 0 \\
    0 & -H_{ij}^*
  \end{pmatrix}
  +
  \delta_{ij}
  \begin{pmatrix}
    0 & \Delta_i
    \\
    \Delta_i & 0
  \end{pmatrix}
  \right]
  \Psi_i
  ,
\end{equation}
where $\Psi_i^\dagger = (c_{i,\up}^\dagger, c_{i,\down})$ and $H_{ij}$
are elements of the normal state Hamiltonian in
Eq.~\eqref{eq:graphene-hopping-hamiltonian}.

We are interested in the transport properties which can be expressed
via the scattering matrix of the system. Following
Ref.~\onlinecite{buttiker:prb:v46:p12485:y1992:scattering_theory}, we
find that the differential conductance matrix
is\cite{blonder:prb:v25:p4515:y1982:BTK}
\begin{equation}
  \label{eq:Gmatrix}
  \begin{split}
    G_{a b}(eV_b)
    &=
    (-1)^{(1-\delta_{a b})}
    \left. \pdiff{I_a}{V_b} \right|_{V_b}
    \\
    &=
    \begin{cases}
      N_b 
      - 
      G_{b b}^\text{ER}
      +
      G_{b b}^\text{DAR}
      ,
      & b = a
      ,
      \\
      G_{a b}^\text{ET}
      -
      G_{a b}^\text{CAR}
      ,
      & a \neq b.
    \end{cases}
    .
  \end{split}
\end{equation}
where $N_b(\varepsilon)$ is the number of propagating modes in lead
$b$ at energy $\varepsilon$, and the conductance matrix elements are
defined in terms of the Nambu space scattering matrix
\begin{equation}
  \mathcal{S}
  =
  \begin{pmatrix}
    \mathcal{S}^{ee} & \mathcal{S}^{eh} \\
    \mathcal{S}^{he} & \mathcal{S}^{hh}
  \end{pmatrix}
\end{equation}
as
\begin{align}
  \label{eq:def:G:ER}
  G_{b b}^\text{ER}
  &=
  \trace{
    \mathcal{S}_{b b}^{e e}(eV_b)
    \mathcal{S}_{b b}^{e e \dagger}(eV_b)
  }
  ,
  \\
  \label{eq:def:G:DAR}
  G_{b b}^\text{DAR}
  &=
  \trace{
    \mathcal{S}_{b b}^{h e}(eV_b)
    \mathcal{S}_{b b}^{h e \dagger}(eV_b)
  }
  ,
  \\
  \label{eq:def:G:ET}
  G_{a b}^\text{ET}
  &=
  \trace{
    \mathcal{S}_{a b}^{e e}(eV_b)
    \mathcal{S}_{a b}^{e e \dagger}(eV_b)
  }
  ,
  \quad (a \neq b)
  \\
  \label{eq:def:G:CAR}
  G_{a b}^\text{CAR}
  &=
  \trace{
    \mathcal{S}_{a b}^{h e}(eV_b)
    \mathcal{S}_{a b}^{h e \dagger}(eV_b)
  }
  .
  \quad (a \neq b)
\end{align}
The conductances in Eqs.~\eqref{eq:def:G:ER} -- \eqref{eq:def:G:CAR}
describe, respectively, local electron reflection (ER), direct Andreev
reflection (DAR), non-local electron transfer (ET), and crossed
Andreev reflection (CAR).

All energies are measured with respect to the equilibrium chemical
potential of the superconductor, and all conductances in this paper
are in units of two times (for spin) the conductance quantum $2
\mathcal{G}_0 = 2 e^2/h$. The current $I_a$ is defined as incoming
from reservoir $a$.

\section{Scattering matrix of a three-terminal device with one
  superconducting terminal}
\label{sec:scatt-matr-multi}

In the following we will study the non-local signal in a
three-terminal device, where terminal 1 is superconducting and
terminals 2 and 3 are normal metals. The non-local
conductance\cite{falci:epl:v54:p255:y2001:correlated_tunneling_superconductor,
  morten:prb:v74:p214510:y2006:circuit_theory_CAR}
\begin{equation}
  G_{3 2}(e V_2)
  =
  G^{\text{ET}}_{3 2}(e V_2)
  -
  G^{\text{CAR}}_{3 2}(e V_2)
\end{equation}
is positive when dominated by ET and negative when dominated by CAR.

We compute $G^{\text{CAR}}_{3 2}$ and $G^{\text{ET}}_{3 2}$ in two
ways: i) $\mathcal{S}^{e e}$ and $\mathcal{S}^{h e}$ are computed
directly in Nambu space using the Hamiltonian \eqref{eq:Nambu}, and
$G^{\text{CAR}}_{3 2}$ and $G^{\text{ET}}_{3 2}$ are found from
\eqref{eq:def:G:ET} and~\eqref{eq:def:G:CAR}. We refer to this as the
\emph{Nambu} approach. ii) We relate $\mathcal{S}^{e e}$ and
$\mathcal{S}^{h e}$ to the scattering matrix $s$ in the normal state
and numerically compute the latter using the Hamiltonian
\eqref{eq:graphene-hopping-hamiltonian}. We call this the
\emph{Normal} approach. Our results using both methods agree, when
applicable. Let us first review how the scattering matrix can be
related to the normal state properties.

Following
Ref.~\onlinecite{beenakker:prb:v46:p124841:y1992:quantumtranspSNmicrojunc},
if the scattering region is well separated from the superconducting
terminal, we can express the scattering matrix $\mathcal{S}$ when
terminal 1 is superconducting in terms of the scattering matrix
\begin{equation}
  \label{eq:Smatr:normal:state}
  s =
  \begin{pmatrix}
    s_{1 1} & s_{1 2} & s_{1 3} \\
    s_{2 1} & s_{2 2} & s_{2 3} \\
    s_{3 1} & s_{3 2} & s_{3 3}
  \end{pmatrix}
  =
  \begin{pmatrix}
    r_{1 1} & t_{1 2} & t_{1 3} \\
    t_{2 1} & r_{2 2} & t_{2 3} \\
    t_{3 1} & t_{3 2} & r_{3 3}
  \end{pmatrix}
\end{equation}
when the whole device is in the normal state. As long as the device is
appreciably smaller than the superconducting coherence length $\xi$,
the normal approach is applicable. For graphene, the induced
superconducting gap is small, $\Delta \sim
\unit{0.1}{\milli\electronvolt}$,\cite{heersche:ssc:v143:p72:y2007:supercondgraphene}
so that the coherence length $\xi$ is on the order of micrometers.

With terminal 1 superconducting, the scattering matrix between the
normal metal terminals $3$ and $2$
is\cite{beenakker:prb:v46:p124841:y1992:quantumtranspSNmicrojunc,
  lesovik:prb:v55:p3146:y1997:nonlinear_NS_scattering_theory}
\begin{align}
  \label{eq:Smatr:ee:general}
  \mathcal{S}^{ee}_{3 2} 
  &=
  t_{3 2} + t_{3 1} \nu^2 \bar{r}_{1 1} M t_{1 2} 
  ,
  \\
  \label{eq:Smatr:he:general}
  \mathcal{S}^{he}_{3 2} 
  &=
  \bar{t}_{3 1} \nu M t_{1 2}
  ,
\end{align}
where the matrix $M$ is
\begin{equation}
  M = \left[ I - \nu^2 r_{1 1} \bar{r}_{1 1} \right]^{-1}.
\end{equation}
The amplitude $\nu$, associated with electron-hole conversion at the
normal-superconducting interface,
is\cite{lesovik:prb:v55:p3146:y1997:nonlinear_NS_scattering_theory}
\begin{equation}
  \label{eq:Smatr:nu} 
  \nu 
  = 
  \frac{\varepsilon}{\Delta} 
  - 
  \signum{\varepsilon} 
  \sqrt{ \left(\frac{\varepsilon}{\Delta} \right)^2 -1 }
  ,
\end{equation}
and the bar ($\bar{g}$) corresponds to time reversal, defined for an
arbitrary quantity $g(\varepsilon)$ as:
\begin{equation}
  \bar{g} = \bar{g}(\varepsilon) = g^*(-\varepsilon).
\end{equation}

The matrix $M$ corresponds to all orders of the process where a hole
emitted from the superconductor returns to the superconductor. At zero
energy, holes propagating with amplitude $\bar{r}_{1 1}$ between
successive interactions with the superconductor retrace exactly the
reverse path of the electrons with amplitude $r_{1 1}$. Thus, at zero
energy holes and electrons do not acquire a phase relative to each
other upon interacting with the scattering region. However, at
non-zero energy, there is a mismatch between the wave vectors of
electron-like and hole-like states, so the time reverse paths
described by the scattering matrices $r_{1 1}$ and $\bar{r}_{1 1}$
will not be exactly opposite to each other. This means that the term
$r_{1 1} \bar{r}_{1 1}$ in $M$ contains many different phases, which
will depend strongly on the disorder configuration. There is therefore
some loss of coherence at non-zero energy due to phase randomization.

For the ET process, described by the scattering matrix element in
Eq.~\eqref{eq:Smatr:ee:general}, there is an interference of two types
of processes: (1) Going directly from $2$ to $3$ without interacting
with the superconductor, and (2) processes involving any number of
electron-hole-electron conversions at the interface to the
superconductor. Similarly, the Andreev process described by
Eq.~\eqref{eq:Smatr:he:general} involves an odd number of
electron-hole conversions at the NS interface.

In the absence of a magnetic field, time-reversal symmetry dictates
\cite{Datta1995}
\begin{equation}
  \bar{s}_{a b}(\varepsilon) 
  = s^*_{a b}(-\varepsilon)
  = s^\dagger_{b a}(-\varepsilon).
\end{equation}
The energy scale of the normal state scattering matrix $s_{a b}$ is
the subband energy, which is determined by the hopping energy $\gamma$
and the width of the ribbon. For the ribbons considered in this paper,
the subband energy is larger than the superconducting pair potential
$\Delta$ by several orders of magnitude. The Fermi energy is
comparable to the subband energies in magnitude. Since, in the regime
we consider, $s_{a b}$ is independent of energy on the scale of
$\Delta$, we write $s_{a b}(\varepsilon) = s_{a b}(0) = s_{a b}$ and
$\bar{s}_{a b} = s_{b a}^\dagger$.

The non-local ET and CAR conductances therefore simplify to
\begin{align}
  \label{eq:G:ET:Smatr}
  \begin{split}
    G^{\text{ET}}_{3 2}
    &=
    \trace{t_{3 2} t_{3 2}^\dagger}
    \\
    &
    \quad
    + 2 \mathrm{Re}
    \nu^2 
    \trace{ t_{1 2} t_{3 2}^\dagger t_{3 1} r_{1 1}^\dagger M }
    \\
    &
    \quad
    +
    \abs{\nu}^4
    \trace{ r_{1 1} t_{3 1}^\dagger t_{3 1} r_{1 1}^\dagger
      M t_{1 2} t_{1 2}^\dagger M^\dagger}
    ,
  \end{split}
  \\
  \label{eq:G:CAR:Smatr}
  \begin{split}
    G^{\text{CAR}}_{3 2}
    &=
    \abs{\nu}^2
    \trace{ t_{1 3} t_{1 3}^\dagger M
      t_{1 2} t_{1 2}^\dagger M^\dagger}
    ,
  \end{split}
\end{align}
where all energy dependence is due to the electron-hole conversion
amplitude $\nu$.

When $\varepsilon \gg \Delta$, Eq.~\eqref{eq:Smatr:nu} gives $\nu \to
0$, and we recover the normal state behavior where only the first
term of Eq.~\eqref{eq:G:ET:Smatr} contributes. However, in the subgap
limit $\varepsilon \ll \Delta$, $\nu \to \im$, and the interaction
with the superconductor contributes. The second term in
Eq.~\eqref{eq:G:ET:Smatr} is due to interference between processes
involving direct transfer of electrons from 2 to 3, and interaction
with the superconducting terminal 1.

In our numerical studies, we calculate the retarded Green's function
matrix $\mathcal{G} = [ E \hat{I} - \hat{H} - \hat{\Sigma}^R ]^{-1}$
and extract the elements $\mathcal{G}_{a b}$ involving the terminals
$a$ and $b$. The calculation of $\mathcal{G}$ uses the recursive
method described in Ref.~\onlinecite{kazymyrenko:prb:v77:p115119}. In
this method, the Green's function of the whole system is found by
adding the sites of the
Hamiltonian~\eqref{eq:graphene-hopping-hamiltonian} to the system one
by one, updating all relevant Green's function elements via the Dyson
equation. The method has the advantage that it can easily be applied
to structures of arbitrary geometry and any number of terminals. After
the Green's function has been found, the scattering matrix
$\mathcal{S}_{a b}$ is extracted via the Fischer-Lee
relations,\cite{fisher:prb:v23:p6851:y1981:fischerlee, Datta1995}
\begin{equation}
  \label{eq:fisher-lee}
  \mathcal{S}_{a b}
  =
  - \mathcal{I}_{a}\delta_{a b}
  + \im \Gamma_{a}^{1/2} \mathcal{G}_{a b} \Gamma_{b}^{1/2}
  .
\end{equation}
Here $\mathcal{I}_{a}$ is the identity matrix (operator) and
$\Gamma_{a} = \im \left( \Sigma_{a} - \Sigma_{a}^\dagger \right)$ is
the linewidth/dephasing matrix which depends on the self energy
$\Sigma_{a}$ of terminal $a$.

\section{Symmetric three-terminal device}
\label{sec:symm-three-term}

The simplest three-terminal device is completely symmetric where the
normal state scattering matrix~\eqref{eq:Smatr:normal:state},
simplifies to
\begin{equation}
  s =
  \begin{pmatrix}
    r & t & t \\
    t & r & t \\
    t & t & r
  \end{pmatrix}
  .
\end{equation}
Unitarity of $s$ gives rise to the relations
\begin{align}
  I &= r r^\dagger + 2 t t^\dagger
  \quad
  \text{and}
  \\
  0 &= t r^\dagger + r t^\dagger + t t^\dagger
  ,
\end{align}
that we make use of in Eqs.~\eqref{eq:G:ET:Smatr}
and~\eqref{eq:G:CAR:Smatr} to find the non-local conductance. We can
express the conductance matrix of such a symmetric device in terms of
the eigenvalues $0 \leq R_n \leq 1$ of the reflection probability
matrix $r r^\dagger$:
\begin{align}
  G^{\text{ET}}_{3 2}
  &=
  \sum_{n}
  \frac{
    (1-R_n)
  }{
    4 (1+R_n)^2
  }
  (3+5 R_n)
  ,
  \\
  G^{\text{CAR}}_{3 2}
  &=
  \sum_{n} \frac{ (1-R_n)^2 }{ 4 (1+R_n)^2 }
  .
\end{align}
It follows from this that the non-local conductance of a symmetric
structure,
\begin{equation}
  \label{eq:evalexpr:G32}
  G_{3 2}
  =
  G^{\text{ET}}_{3 2}
  -
  G^{\text{CAR}}_{3 2}
  =
  \sum_{n}
  \frac{(1-R_n)}{ 2 (1+R_n)^2 }
  (1+3 R_n)
  ,
\end{equation}
is always ET dominated (positive). By a completely analogous
calculation we also find that the local conductance,
\begin{equation}
  \label{eq:evalexpr:G22}
  G_{2 2}
  =
  \sum_{n}
  \frac{ (1-R_n) }{ 2 (1+R_n)^2 }
  \left[
    3 + R_n (2-R_n)
  \right]
  ,
\end{equation}
is naturally also positive.

A few simple conclusions can be drawn from these expressions. First,
when the device is perfectly transparent for the $N_\varepsilon$
contributing modes at the Fermi energy, the contributing modes have
$R_n = 0$, the others have $R_n=1$. The local and non-local
conductances become,
\begin{align}
  \label{eq:evalexpr:transparent:local}
  G_{2 2}
  &= 
  \frac{3}{2} N_{2}
  ,
  \\
  \label{eq:evalexpr:transparent:non-local}
  G_{3 2}
  &=  
  \frac{1}{2} N_{2}
  ,
\end{align}
where $N_2$ is the number of modes contributing to the current at the
Fermi energy. These results have the following simple explanation:
When the voltage is raised in terminal 2, $N_2$ conducting modes are
injected into the structure via this terminal. Since the device is
symmetric, half of these modes go directly to terminal 3, producing a
current $N_2/2$ in this terminal. The other half of the $N_2$ incoming
modes interact with the superconductor at terminal 1, and are Andreev
reflected back to terminal 2 as holes. These modes contribute $2 \cdot
(N_2/2)$ to the current in terminal 2. The total current in terminal 2
is therefore $(2 + 1) (N_2/2) = 3/2 N_2$. At low bias the Andreev
reflected holes retrace exactly the trajectory of the incoming
electrons because of time-reversal symmetry, and they therefore only
contribute to the current in terminal 2.

When all the terminals are connected to the central device with tunnel
contacts, we can expand the local and non-local conductances in the
$\delta = (1-R_n)/2 \ll 1$ for the contributing modes. We find that
\begin{align}
  G_{2 2}
  &= 
  N_{2} \delta
  + \order{\delta}{2}
  ,
  \\
  G_{3 2}
  &= 
  N_{2} \delta
  + \order{\delta}{2}
  ,
\end{align}
so we recover the normal state results, where both the local and
non-local signals vanish linearly with $\delta$. Transport between
terminals 2 and 3 involving the superconductor involves higher orders
in $\delta$ and does therefore not contribute in this limit.

\begin{figure}
  \centering
  \includegraphics[width=\columnwidth]{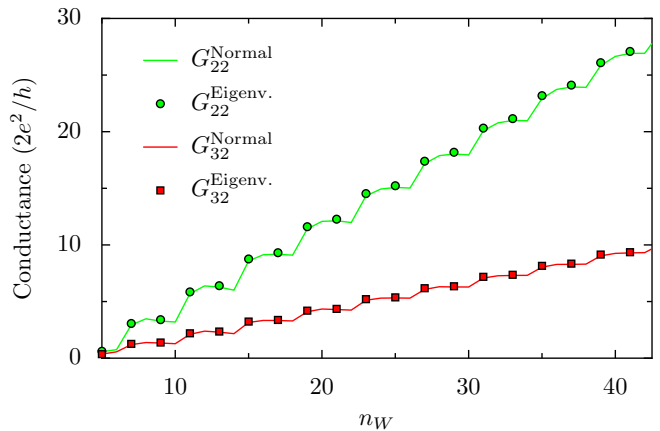}
  \caption{Local and non-local conductance of a symmetric device as a
    function of the ribbon width $n_W$. Comparing calculations done
    directly from the scattering matrix (lines) with calculations
    using the eigenvalues of the reflection matrix (symbols), we see
    that the two methods give identical results.}
  \label{fig:evalcomparison}
\end{figure}

We check the consistency between the eigenvalue
expressions~\eqref{eq:evalexpr:G32} and~\eqref{eq:evalexpr:G22} and
our numerical routines by calculating the local and non-local
conductances $G_{2 2}$ and $G_{3 2}$ in a symmetric three-terminal
graphene device, as explained further in
Sec.~\ref{sec:graphene-three-term}. As can be seen from
Fig.~\ref{fig:evalcomparison}, where the conductances are plotted as a
function of the width $n_W$ of the nanoribbons, we have excellent
agreement between the eigenvalue expressions (symbols) and the results
found directly from Eq.~\eqref{eq:Gmatrix} (lines). Note that
Eq.~\eqref{eq:Gmatrix} is valid for any width $n_W$, while only even
$n_W$ give a truly symmetric device when built from zigzag
nanoribbons, so only such data points are shown. However, we find that
the results found from (ab-)using the eigenvalue expressions when
$n_W$ is odd are also very close to the numerical results. This is not
surprising, since as long as many modes contribute to the current,
small alterations of the geometry should not have a big impact on the
total current.

\section{Asymmetric three-terminal device}
\label{sec:graphene-three-term}

\begin{figure}
  \centering
  \includegraphics[width=0.8\columnwidth]{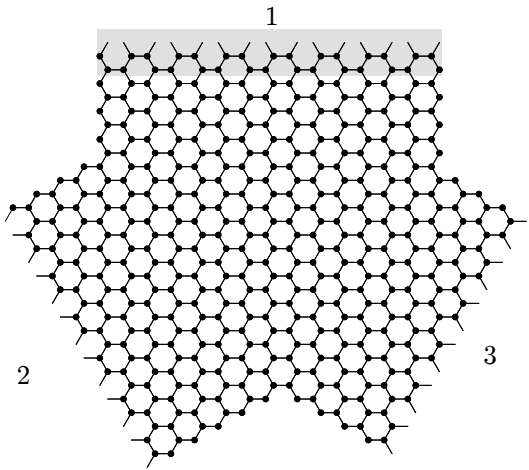}
  \caption{Three-terminal graphene fork consisting of three
    semi-infinite graphene zigzag nanoribbons connected together. The
    top lead is allowed to become superconducting by inserting a
    non-zero pair potential $\Delta$ in this region, according to the
    Hamiltonian in~\protect\eqref{eq:Nambu}.}
  \label{fig:graphene:three:terminal:device}
\end{figure}

Having found that ET dominates non-local transport in a symmetric
device, we turn our investigation to asymmetric devices. We do this
numerically, by calculating the scattering matrix in a three-terminal
device obtained by joining three semi-infinite zigzag graphene
nanoribbons as shown in Fig.~\ref{fig:graphene:three:terminal:device}.

The width of a zigzag graphene nanoribbon, $W = \sqrt{3} a n_W/2$, is
determined by the minimal number of bonds $n_W$ that must be broken to
cut the
ribbon.\cite{fujita:jpsj:v65:p1920:y1996:localized_zigzag_edge_states}
$a$ is the lattice constant of graphene, $a =
\unit{2.46}{\angstrom}$.\cite{wallace:pr:v71:p622:y1947}

For a wide ribbon, $n_W \gg 1$, the energy of the $m$'th transverse
subband is to a good
approximation\cite{rycerz:npys:v3:p172:y2007:valleyfilter}
\begin{equation}
  E_m
  =
  \left( m + \frac{1}{2} \right)
  \frac{\pi \gamma}{n_W}
  ,
  \qquad m = 1,2,\ldots,
\end{equation}
where $\gamma$ is the nearest neighbor hopping energy on the graphene
lattice. The superconducting coherence length
\begin{equation}
  \xi
  = 
  \frac{\hbar v_F}{\Delta} 
  =
  \frac{\sqrt{3}}{2}
  \frac{\gamma}{\Delta} 
  a
\end{equation}
will typically be of the order of micrometers, so the normal approach
should be applicable for nanoribbons up to $\unit{1}{\micro\metre}$
wide, or $n_W \sim 10^4$.

\subsection{Consistency checks}

\begin{figure}
  \centering
  \includegraphics[width=\columnwidth]{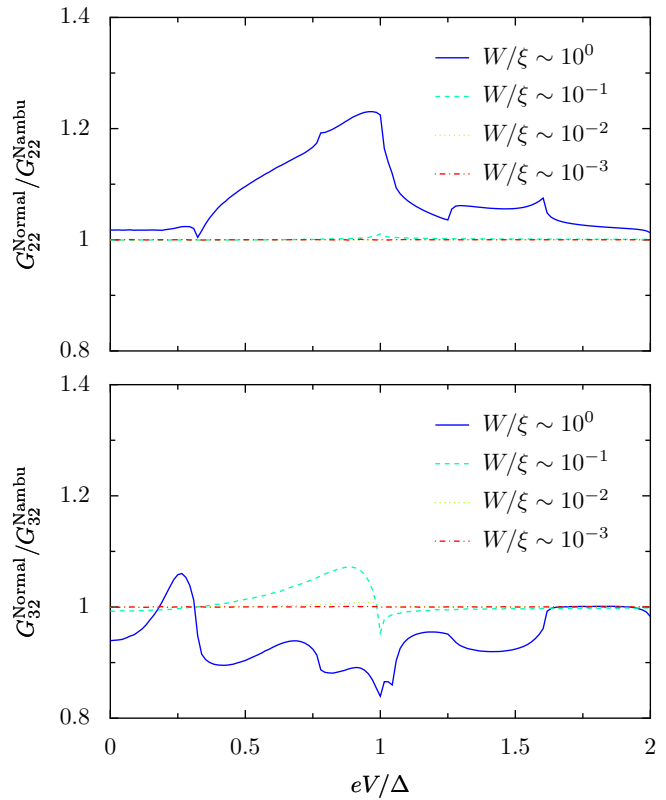}
  \caption{ Ratio of conductance calculated with the normal method
    (extracted from the normal state scattering matrix) to conductance
    calculated directly with the Nambu space Hamiltonian. When $W \ll
    \xi$ the two methods give identical results. The situation
    considered in this paper corresponds to $W/\xi \sim
    10^{-4}$. Upper (lower) panel: local (non-local) conductance.}
  \label{fig:method:comparison}
\end{figure}

In Fig.~\ref{fig:method:comparison} we compare the conductance
extracted from the normal state scattering matrix $s$ to that found by
direct evaluation of the full scattering matrix $S$ in Nambu space.
The calculations are done for a device of the type shown in
Fig.~\ref{fig:graphene:three:terminal:device}. The leads are all
semi-infinite zigzag ribbons, and we set $\Delta = 0$ everywhere
except in terminal 1 (shaded area in
Fig.~\ref{fig:graphene:three:terminal:device}). As can be seen from
Fig.~\ref{fig:method:comparison}, where we show the ratio between the
conductance calculated with the two methods,
$G_{ij}^\text{Normal}/G_{ij}^\text{Nambu}$, the agreement between the
two methods is excellent as long as $W/\xi \ll 10^{-1}$, where $W$ is
the width of the ribbons and $\xi$ is the superconducting coherence
length.
 
\begin{figure}
  \centering
  \includegraphics[width=0.8\columnwidth]{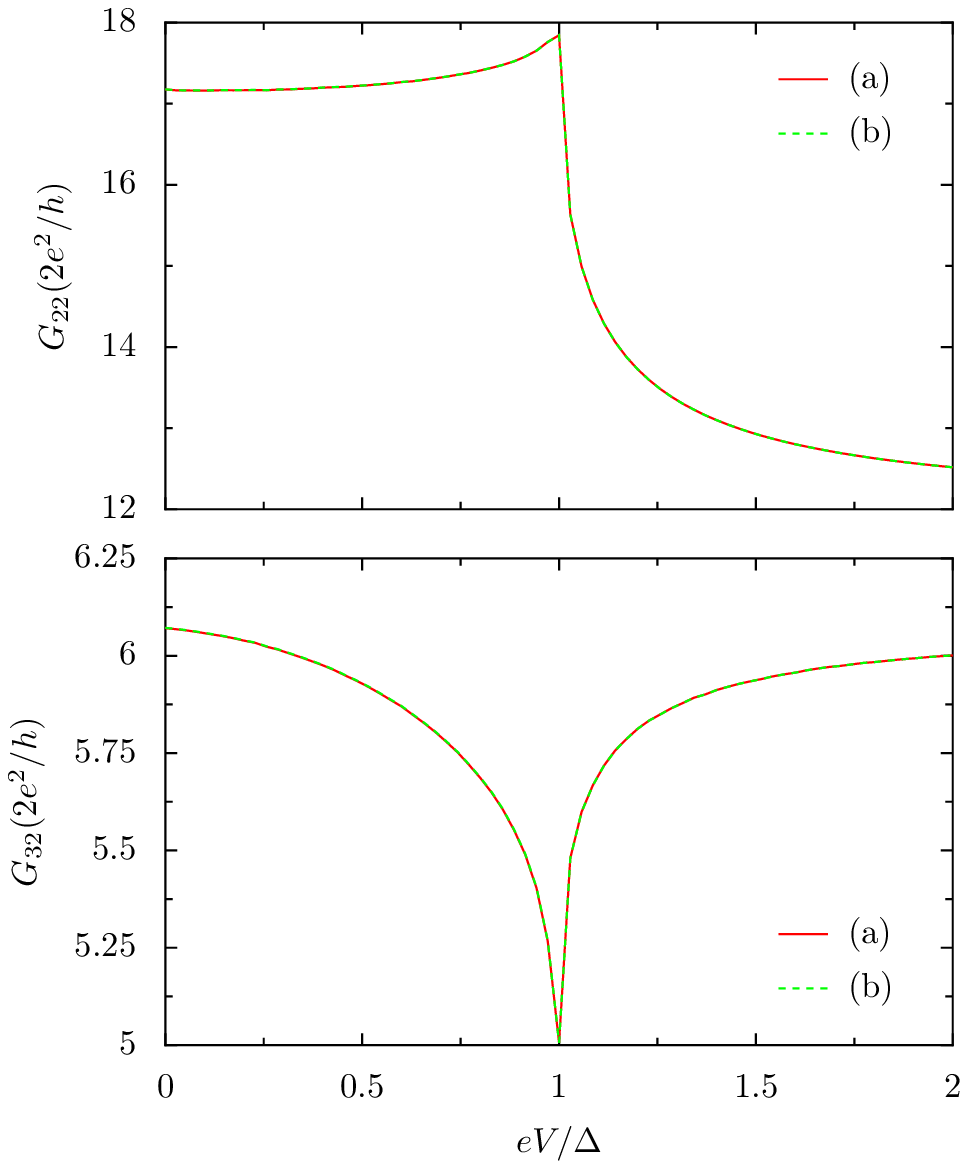}
  \subfloat[]{
    \label{fig:sucond:whole:device}
    \includegraphics[width=0.4\columnwidth]{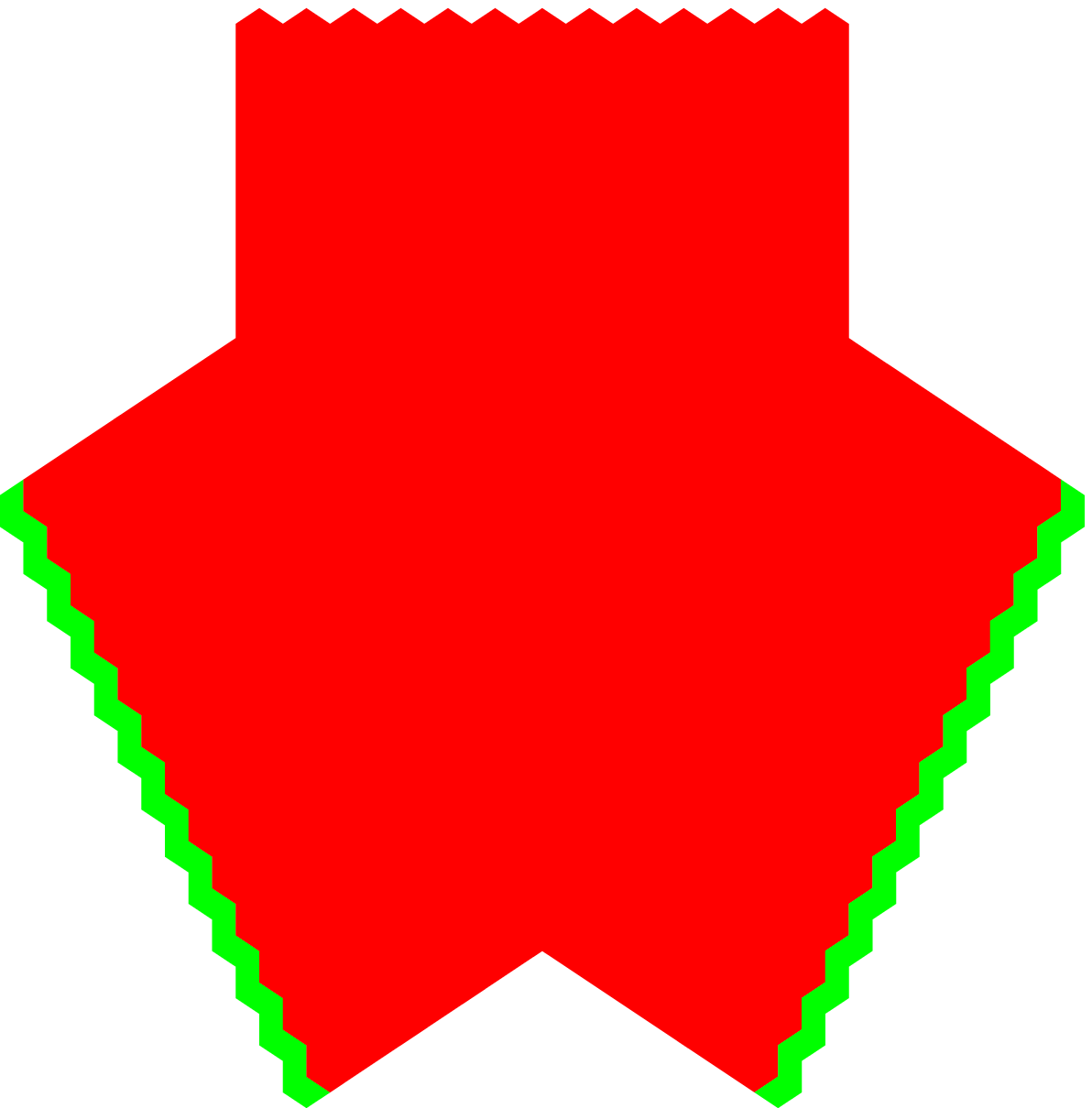}
  }
  \subfloat[]{
    \label{fig:sucond:top:lead}
    \includegraphics[width=0.4\columnwidth]{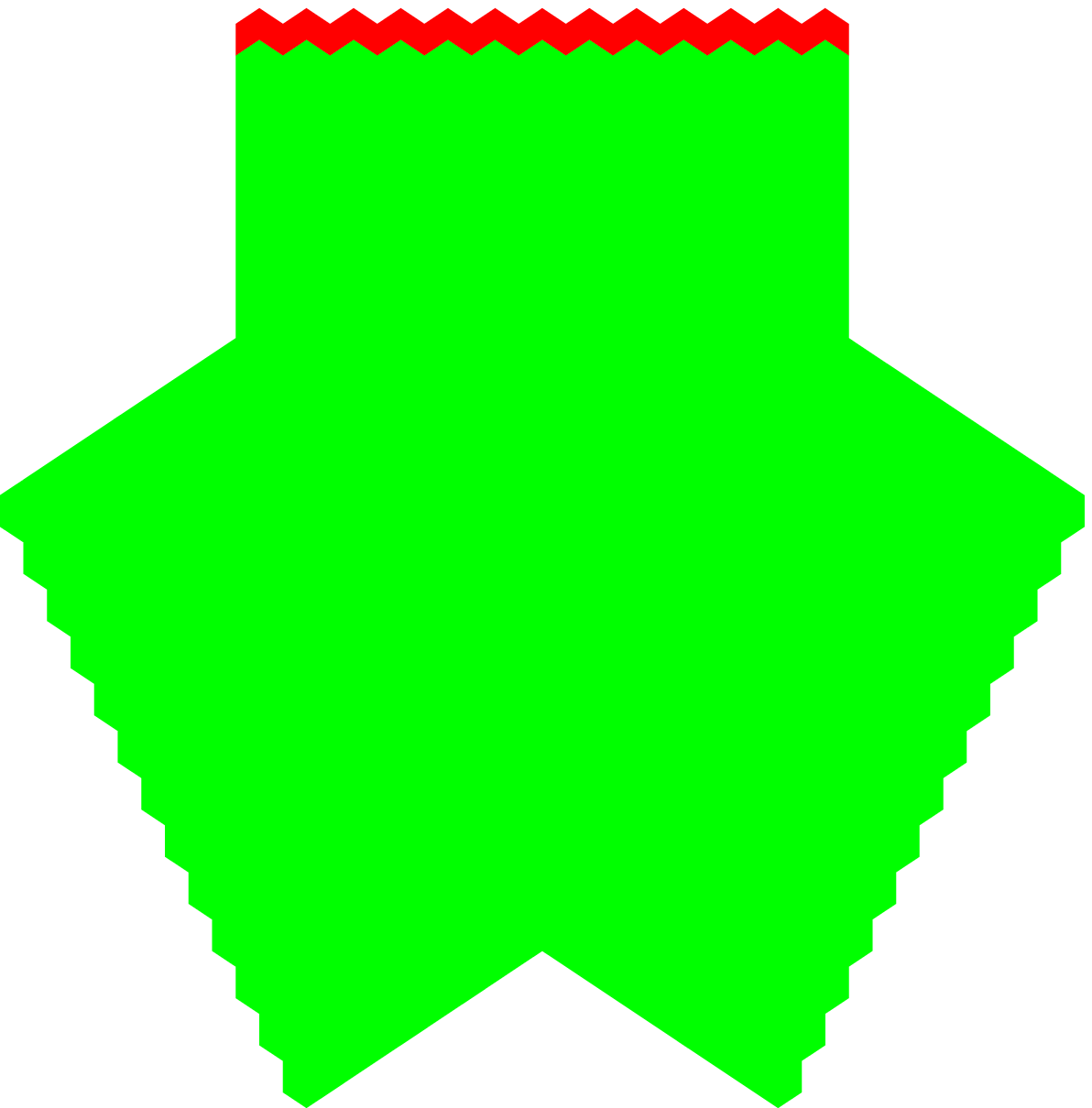}
  }
  \caption{\label{fig:sucond:posdep} Dependence of local and non-local
    conductance on the position of the normal-superconductor (NS)
    interface. In~\protect\subref{fig:sucond:whole:device}, $\Delta$
    is non-zero in the whole scattering region, while
    in~\protect\subref{fig:sucond:top:lead} $\Delta$ is non-zero only
    in the top terminal. The conductances calculated for the two cases
    are identical, implying that the exact position of the NS
    interface is not important.}
\end{figure}

Also, since $\xi \gg W$, the exact position of the boundary between
the normal ($\Delta = 0$) and superconducting ($\Delta \neq 0$)
regions does not influence our results. This can be seen explicitly
from Figs.~\ref{fig:sucond:whole:device}
and~\ref{fig:sucond:top:lead}, where we compare the conductance
matrices for systems when the scattering region is, respectively,
entirely mixed with, or separated from, the superconducting
region. There is no dependence on the exact position of the NS
interface, as should be expected.

\subsection{Varying the width of the superconductor}
\label{sec:vary-width-superc}

\begin{figure}
  \centering
  \subfloat[Equal doping the whole device, $E_F = E_F^\text{leads} =
  0.9\gamma$.]{
    \label{fig:asymmetricdevice:varytop:equal:doping}
    \includegraphics[width=\columnwidth]{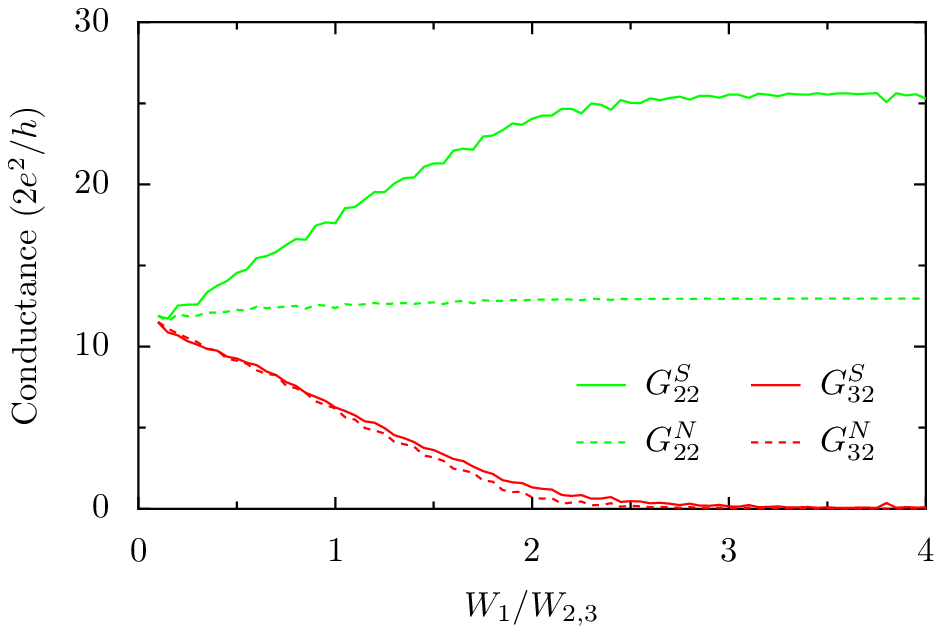}
  }
  \\
  \subfloat[Increased doping in terminals, $E_F = 0.9\gamma,
  E_F^\text{leads} = 1.1\gamma$.]{
    \label{fig:asymmetricdevice:varytop:different:doping}
    \includegraphics[width=\columnwidth]{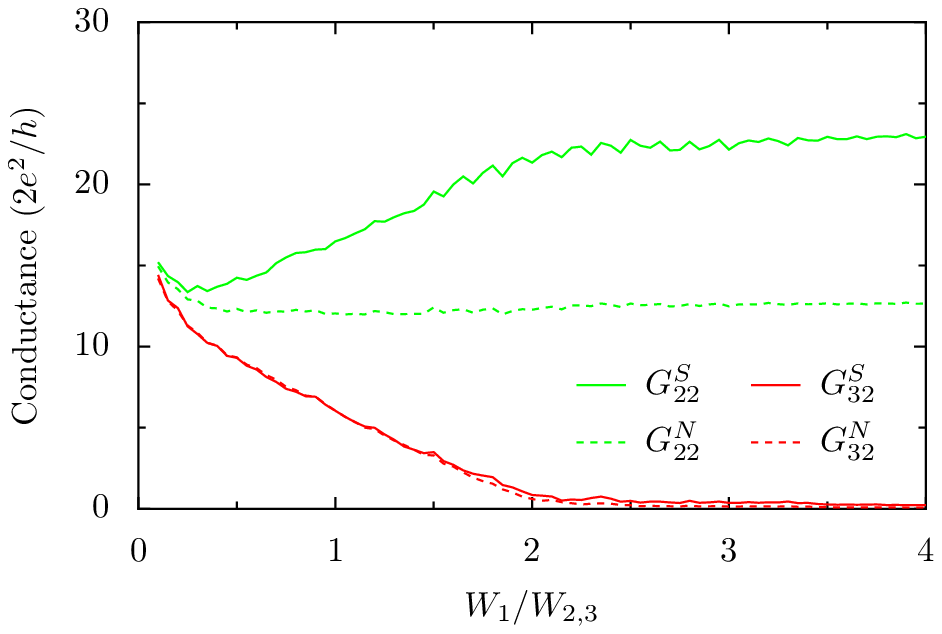}
  }
  \caption{Local and non-local conductances as functions of the width
    $W_1$ of the superconducting lead. $W_2 = W_3$ is kept fixed
    ($n_{W_{2,3}} = 20$). For comparison we show the conductance both
    when the top lead is in the normal state (dashed) and when it is
    superconducting (solid).}
  \label{fig:asymmetricdevice:varytop}
\end{figure}

We now turn to the numerical calculations of the local and non-local
conductances for an asymmetric device. We first vary the width $W_{1}$
of the superconducting lead, keeping the width of the normal terminals
fixed, $W_{2} = W_{3}$. The doping throughout the device is set to a
high value to ensure that many modes contribute to the transport. The
conductances involving the normal terminals are calculated when the
top terminal is superconducting (superscript $S$), and compared for
reference to the conductance when the whole device is in the normal
state (superscript $N$). As seen in
Fig.~\ref{fig:asymmetricdevice:varytop}, when the superconducting lead
is very narrow, there is very little coupling via the induced
superconducting order parameter, so $G_{2 2}^{S} \approx G_{2
  2}^{N}$. However, as the width of the superconducting lead
increases, more of the incoming quasiparticles are coupled via the
induced superconducting order parameter, and the local conductance
asymptotically approaches twice the normal state conductance, $G_{2
  2}^S \approx 2 G_{2 2}^{N}$.  In other words, the interaction with
the superconductor essentially involves all the incoming modes, which
are therefore Andreev reflected. This resembles the situation in a
strongly coupled two-terminal NS
junction.\cite{beenakker:prb:v46:p124841:y1992:quantumtranspSNmicrojunc}
From Fig.~\ref{fig:asymmetricdevice:varytop} we see that the
contribution of Andreev reflection to the local conductance has
reached its asymptotic value when $W_{1}/W_{2,3} \approx 2$.

The picture is essentially unchanged if we allow for different doping
levels in the central device and the terminals (see
Fig.~\ref{fig:asymmetricdevice:varytop:different:doping}), as long as
the number of contributing modes in the central region is still large.

Contrary to what is seen for the local conductance, the non-local
conductance $G_{32}$ is only weakly influenced by the presence of the
superconductor. Except for a slight enhancement of the signal around
$W_{1}/W_{2,3} \approx 2$, the non-local conductance remains
essentially unchanged when turning on the superconductor. This is in
accordance with what was found for the symmetric device in
Sec.~\ref{sec:symm-three-term}, namely that the Andreev reflected
holes retrace the path of the incoming electrons, giving a negligible
contribution to the non-local conductance. Again, different doping
levels in the central device and the terminals do not change the
picture qualitatively, as seen from
Fig.~\ref{fig:asymmetricdevice:varytop:different:doping}.

\subsection{Varying the width of normal terminal 3}
\label{sec:varying-width-normal}

\begin{figure}
  \centering
  \subfloat[Equal doping the whole device, $E_F = E_F^\text{leads} =
  0.9\gamma$.]{
    \label{fig:asymmetricdevice:varyright:equal:doping}
    \includegraphics[width=\columnwidth]{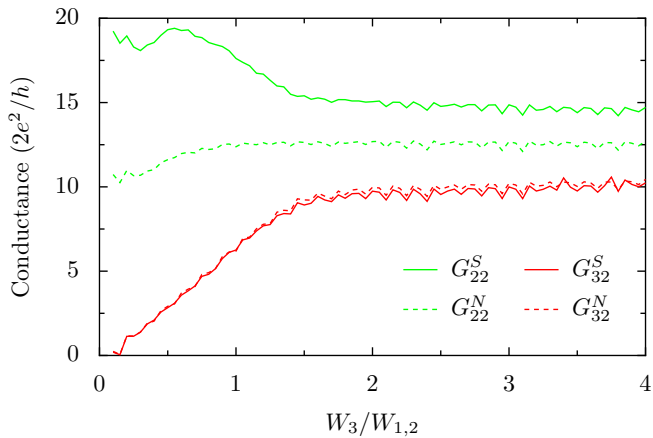}
  }
  \\
  \subfloat[Increased doping in terminals, $E_F = 0.9\gamma,
  E_F^\text{leads} = 1.1\gamma$.]{
    \label{fig:asymmetricdevice:varyright:different:doping}
    \includegraphics[width=\columnwidth]{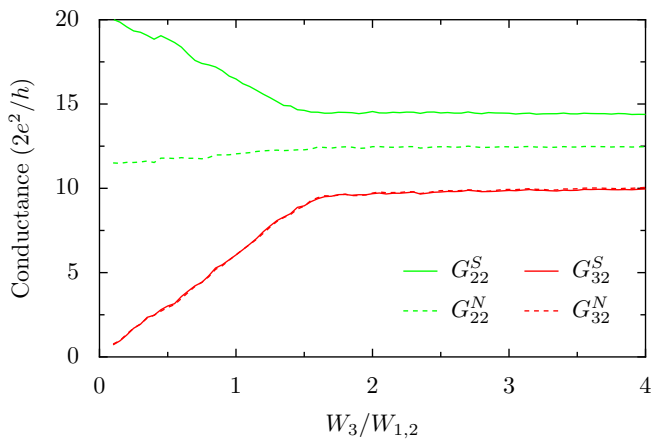}
  }
  \caption{ Local and non-local conductance as a function of the width
    $W_3$ of the right normal lead. $W_1 = W_2$ is kept fixed
    ($n_{W_{1,2}} = 20$). Doping levels and numerical method is the
    same as in Fig.~\protect\ref{fig:asymmetricdevice:varytop}.}
  \label{fig:asymmetricdevice:varyright}
\end{figure}

We also vary the width $W_{3}$ of the normal terminal 3, keeping the
widths of the voltage terminal 2 and the superconducting terminal 1
fixed. As can be seen from Fig.~\ref{fig:asymmetricdevice:varyright}
the non-local conductance, $G_{32}$, is (nearly) zero when terminal 3
is very narrow. This is natural, since the subband energies in
terminal 3 increase as the terminal is made narrower, hindering any
modes from carrying current at the Fermi energy. As the normal lead 3
widens, more and more channels in this lead are opened, and we get an
increase in the conductance due to the opening of new subbands. The
current in terminal 3 saturates when all available subbands
participate in the transport. We observe that the superconductor has
very little influence on the non-local conductance. As in
Sec.~\ref{sec:vary-width-superc}, these results persist if we allow
for different doping levels in the central device and the terminals,
demonstrating the robustness of the results.

In contrast to the non-local conductance $G_{3 2}$, the local
conductance $G_{2 2}$ is strongly affected by the
superconductor. $G_{2 2}^{S}$ doubles compared to the normal state
value $G_{2 2}^{N}$ when the other normal lead becomes vanishingly
small, $W_{3}/W_{1,2} \to 0$. This is as expected, since we are
effectively left with a strongly coupled two-terminal NS junction
involving only normal terminal 2 and superconducting terminal 1. In
the opposite limit, when $W_3/W_{1,2} \gg 1$, the ratio
$G_{22}^{S}/G_{22}^N$ approaches an asymptotic value due to the fact
that only a certain fraction of the finite number $N_2$ of incoming
modes still interact with the superconductor, although the majority
of these incoming modes are transferred directly to terminal 3 in this
limit.

\section{Non-local conductance dominated by CAR}
\label{sec:car-dominance}

\subsection{Zero modes of zigzag nanoribbons}

A zigzag graphene ribbon supports special current carrying zero energy
modes. When the doping is low (close to the Dirac point), the density
of the zero energy modes is localized along the ribbon
edges,\cite{fujita:jpsj:v65:p1920:y1996:localized_zigzag_edge_states,
  nakada:prb:v54:p17954:y1996:graphen_ribbon_edges} while the current
is distributed approximately uniformly across the width of the
ribbon.\cite{munos-rojas:prb:v74:p195417:y2006:zigzag_current_homogeneous,
  zarbo:epl:v80:p47001:y2007:spatial_distribution_current_ribbons,
  nakakura:jpsj:v78:p065003:y2009:uniform_current_zigzag_ribbon}
Associated with the zero energy modes is a quantum number called
pseudoparity, arising from the fact that the unit cell of the
honeycomb lattice contains two
atoms.\cite{akhmerov:prb:v77:p205416:y2008:valley_valve,
  cresti:prb:v77:p233402:y2008:valley_valve,
  rainis:prb:v79:p115131:y2009:AR_in_graphene_ribbons} The
conservation of pseudoparity in a zigzag ribbon has dramatic
consequences for the transport in a normal-superconducting (NS)
junction when only the zero modes contribute, i.e. for Fermi energies
below the first subband energy $E_1$. In this regime each lead only
supports one incoming and one outgoing mode. These two modes have
either the same or opposite pseudoparities, depending on whether $n_W$
is even or odd,
respectively.\cite{rainis:prb:v79:p115131:y2009:AR_in_graphene_ribbons}
As a consequence of this, in a zigzag ribbon NS junction, either
normal reflection or direct Andreev reflection will be entirely
suppressed, depending on the value of $n_W$ (modulo
2).\cite{rainis:prb:v79:p115131:y2009:AR_in_graphene_ribbons} In a
three-terminal device, pseudoparity is not a good quantum number, but
when the transport involves only the zero modes, traces of even/odd
effects can still be seen in the contributions due to Andreev
reflection.

\subsection{CAR dominance}

\begin{figure}
  \centering
  \includegraphics[width=0.8\columnwidth]{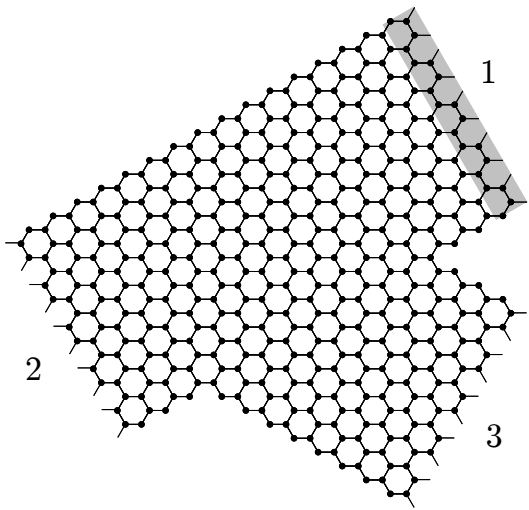}
  \caption{Zigzag ribbon with a third terminal at $120\degree$, where
    the top lead (shaded) be superconducting. The depicted structure
    corresponds to $n_W^\text{ribbon} = n_W^\text{leg} = 10$.}
  \label{fig:ribbonwleg:system}
\end{figure}

Motivated by the results of Refs.~\onlinecite{
  rycerz:arxiv:0709.3397:y2007:aharonov_bohm,
  iyengar:prb:v78:p235411:y2008,
  katsnelson:prb:v78:p075417:y2008:evanescent_transport_graphene_dots},
where it was found that a $120\degree$ kink in a graphene ribbon can
in certain situations completely block the electron flow, we construct
a device as depicted in Fig.~\ref{fig:ribbonwleg:system}, consisting
of a zigzag ribbon with a third terminal protruding at an angle of
$120\degree$. The top terminal (1) is superconducting, while the left
(2) and lower right (3) terminals are normal. We define
$n_W^\text{ribbon} = n_W^1 = n_W^2$ and $n_W^\text{leg} = n_W^3$ and
set $n_W^\text{ribbon} = n_W^\text{leg} = n_W$ in this section. The
superconductor is heavily doped, while the doping in the
non-superconducting part of the structure is kept close to the Dirac
point. We study the transport properties as a function of back gate
voltage $V_g$, which specifies the overall doping of the device, in
the regime where only the zero modes contribute in the normal part of
the device, $\abs{V_g} < E_1$.

\begin{figure}
  \centering
  \includegraphics[width=\columnwidth]{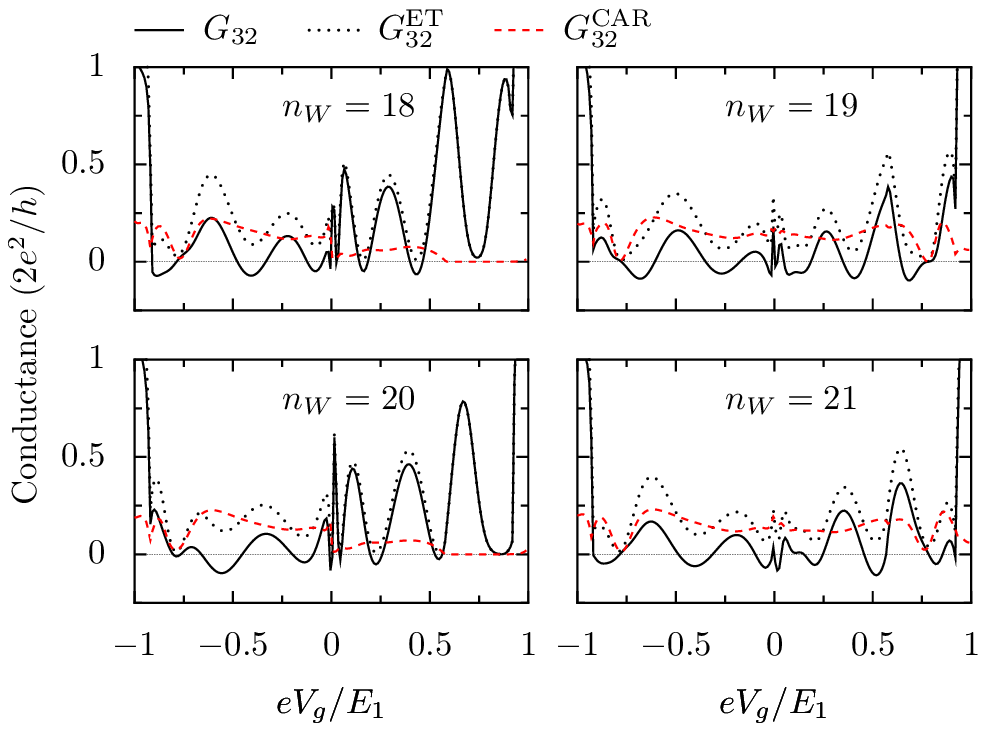}
  \caption{Non-local conductance $G_{32}$ (solid black line) at zero
    bias as a function of back gate voltage $V_g$. The conductance
    changes sign due to the competition between ET (dotted black line)
    and CAR (dashed red line). A negative $G_{32}$ corresponds to CAR
    dominating ET. The second subband starts to contribute when
    $\abs{eV_g} > E_1$.}
  \label{fig:ribbonwleg:oddeven:group2}
\end{figure}

\begin{figure}
  \centering
  \includegraphics[width=\columnwidth]{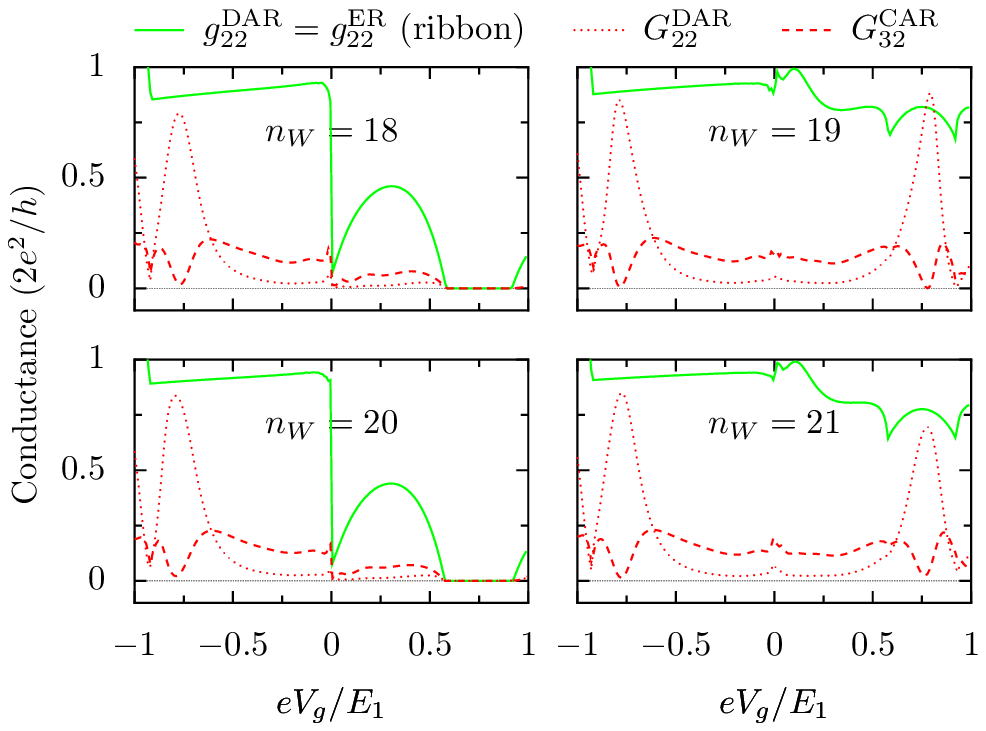}
  \caption{Comparison of Andreev reflection in a two terminal NS
    ribbon and in a three-terminal device as that shown in
    Fig.~\protect~\ref{fig:ribbonwleg:system}. The general behavior of
    direct Andreev reflection in the NS ribbon, $g_{22}^\text{DAR}$
    (solid green line), is reflected in the three-terminal crossed and
    direct Andreev reflection conductance $G_{32}^\text{CAR}$ (dashed
    red line) and $G_{22}^\text{DAR}$ (dotted red line).}
  \label{fig:ribbonwleg:oddeven:group2:tuncomp}
\end{figure}

The numerical results in Fig.~\ref{fig:ribbonwleg:oddeven:group2} show
that the zero bias non-local conductance $G_{32}$ changes sign several
times in the regime $\abs{eV_g} < E_1$. This demonstrates that CAR can
in principle dominate ET in rather specific geometries. The non-local
conductance changes sign due to close competition between ET and
CAR. The oscillatory pattern is determined by the distance between the
superconducting terminal and the scattering centre at the junction
between the ribbon and terminal 3. The contribution from Andreev
reflection is insensitive to this length, as seen from
Fig.~\ref{fig:ribbonwleg:oddeven:group2}. Also, we observe that when
$\abs{eV_g} > E_1$, a new subband starts contributing to ET, while CAR
remains approximately unchanged.

Finally, we also make a short comment on the even/odd behavour of our
three-terminal device. By comparing our results with the conductance
$g_{22} = 2 g_{22}^\text{DAR}$ in a two terminal NS ribbon (similar to
the device in Fig.~\ref{fig:ribbonwleg:system}, but without terminal
3), we see that the even/odd behaviour of $g_{22}^\text{DAR}$ is
reflected in $G_{32}^\text{CAR}$ and $G_{22}^\text{DAR}$, as can be
seen from Fig.~\ref{fig:ribbonwleg:oddeven:group2:tuncomp}. According
to the results of
Ref.~\onlinecite{rainis:prb:v79:p115131:y2009:AR_in_graphene_ribbons},
incoming carriers at positive and negative $V_g$ have opposite
(identical) pseudoparities in a ribbon with $n_W$ even (odd). With our
chosen doping in the superconductor, this leads to a blocking of
Andreev reflection for positive $V_g$ in the two-terminal ribbon
(solid green line) when $n_W$ is even. This feature is still manifest
in the local DAR (dotted red line) and non-local CAR (dashed red line)
contributions to the conductance in the three-terminal device.

\section{Conclusion}
\label{sec:conclusion}

In this work we have studied the contribution from CAR and ET to the
non-local transport in a devices having two normal metal terminals and
one superconducting terminal.

ET dominates CAR in a symmetric three-terminal device when the
superconducting coherence length $\xi$ greatly exceeds the device
dimensions. The Andreev conversion process then contributes almost
exclusively to direct Andreev reflection, due to vanishing wave vector
mismatch between electrons and back-reflected holes. This regime is
relevant for ballistic transport in graphene nanoribbons devices of
dimensions up to the micrometer scale.  Superconductivity can be
induced in such structures via the proximity effect.

For most asymmetric systems ET dominates the non-local
conductance. However, for asymmetric devices where the direct ET
contribution can be suppressed, marginal CAR dominated charge
transport is possible. The crossover from CAR to ET dominated
transport in such a device can be induced by varying the overall
doping of the device via a back gate.

\begin{acknowledgments}
  This work was supported by the Research Council of Norway through
  grant no. 167498/V30.
\end{acknowledgments}


\bibliography{graphenefork}

\end{document}